\documentclass[10pt,aps,prx,twocolumn,nofootinbib,superscriptaddress]{revtex4-2}
\usepackage[utf8]{inputenc}
\usepackage[english]{babel}
\usepackage[caption=false, position=top]{subfig}
\usepackage{braket}
\usepackage{amsmath}
\usepackage{siunitx}
\usepackage[breaklinks=true,colorlinks,citecolor=blue,linkcolor=blue,urlcolor=blue]{hyperref}
\usepackage{orcidlink}
\usepackage{txfonts}

\newenvironment{wide}{\onecolumngrid}{\twocolumngrid\noindent}
\newcommand\norm[1]{\left\lVert#1\right\rVert}
\newcommand{\trace}[1]{\mathrm{tr}[#1]}
\newcommand{\figlabel}[1]{Fig.~\ref{#1}}
\newcommand{\seclabel}[1]{Section~\ref{#1}}
\newcommand{\applabel}[1]{Appendix~\ref{#1}}
\newcommand{\hc}{\text{h.c.}}

\makeatletter
\def\bibsection{%
   \par
   \begingroup
    \baselineskip26\p@
    \bib@device{\hsize}{72\p@}%
   \endgroup
   \nobreak\@nobreaktrue
   \addvspace{19\p@}%
  }%
\makeatother

\begin{document}

\title{Optimizing edge-state transfer in a Su-Schrieffer-Heeger chain via hybrid analog-digital strategies}
\author{Sebastián V. Romero$^{\orcidlink{0000-0002-4675-4452}}$}
\affiliation{TECNALIA, Basque Research and Technology Alliance (BRTA), 48160 Derio, Spain}
\affiliation{Department of Physical Chemistry, University of the Basque Country (UPV/EHU), Apartado 644, 48080 Bilbao, Spain}
\author{Xi Chen$^{\orcidlink{0000-0003-4221-4288}}$}
\affiliation{Department of Physical Chemistry, University of the Basque Country (UPV/EHU), Apartado 644, 48080 Bilbao, Spain}
\affiliation{EHU Quantum Center, University of the Basque Country (UPV/EHU), Apartado 644, 48080 Bilbao, Spain}
\author{Gloria Platero$^{\orcidlink{0000-0001-8610-0675}}$}
\affiliation{Instituto de Ciencia de Materiales de Madrid (CSIC), Cantoblanco, E-28049 Madrid, Spain}
\author{Yue Ban$^{\orcidlink{0000-0003-1764-4470}}$}
\email[]{ybanxc@gmail.com}
\affiliation{Departamento de F\'isica, Universidad Carlos III de Madrid, Avda. de la Universidad 30, 28911 Legan\'es, Spain}
\affiliation{TECNALIA, Basque Research and Technology Alliance (BRTA), 48160 Derio, Spain}
\date{\today}

\begin{abstract}
The Su-Schrieffer-Heeger (SSH) chain, which serves as a paradigmatic model for comprehending topological phases and their associated edge states, plays an essential role in advancing our understanding of quantum materials and quantum information processing and technology. In this paper, we introduce a hybrid analog-digital protocol designed for the nonadiabatic yet high-fidelity transfer of edge states in an SSH chain, featuring two sublattices, A and B. The core of our approach lies in harnessing the approximate time-dependent counterdiabatic (CD) interaction, derived from adiabatic gauge potentials. However, to enhance transfer fidelity, particularly in long-distance chains, higher-order nested commutators become crucial. To simplify the experimental implementation and navigate computational complexities, we identify the next-to-nearest-neighbor hopping terms between sublattice A sites as dominant CD driving and further optimize them by using variational quantum circuits. Through digital quantum simulation, our protocol showcases the capability to achieve rapid and robust solutions, even in the presence of disorder. This analog-digital transfer protocol, an extension of quantum control methodology, establishes a robust framework for edge-state transfer. Importantly, the optimal CD driving identified can be seamlessly implemented across various quantum registers, highlighting the versatility of our approach.
\end{abstract}

\maketitle

\section{Introduction}

Building on Feynman's seminal 1982 conjecture \cite{feynman2018simulating}, quantum simulation has emerged as a valuable toolkit, allowing the study of complex form of behavior in condensed matter systems by leveraging easily controllable quantum systems \cite{quantum-simulation}.
In the realm of quantum computing platforms, topological quantum systems have garnered substantial attention \cite{topological-QC} due to their inherent robustness, which arises from topological protection mechanisms.
Among them, the Su-Schrieffer-Heeger (SSH) model \cite{su1979solitons,SSH1,SSH2,SSH3}, which serves as a tight-binding model for the simplest one-dimensional (1D) lattice, encompasses topologically safeguarded edge states. This model has sparked exploration of topological properties across diverse quantum registers, known for their high tunability, including superconducting Xmon qubit chains \cite{mei2018robust}, a 1D bichromatic lattice \cite{Atala2013-lattice}, and arrays of optically trapped Rydberg atoms \cite{ssh-Rydberg}, as well as semiconductor quantum dots \cite{SSH-QDs}.

Digital quantum simulation (DQS) offers a remarkable avenue for investigating a diverse range of intricate models, without the need for direct laboratory engineering. Versatile digital-simulation methodologies have been harnessed to explore various domains, including topological phases \cite{Digital-topological1, Digital-topological2, Digital-topological3} and quench dynamics \cite{Digital-quench}.
Unlike analog quantum simulation, which is confined to specific quantum system types, DQS stands out for its execution via sequences of quantum logic gates. This can be achieved using present-day quantum simulators and holds promise for future fault-tolerant quantum computers \cite{preskill2018quantumcomputingin}. To foster the development of DQS, hybrid quantum-classical optimization algorithms such as variational quantum algorithms \cite{cerezo2021variational} grounded in gradient descent \cite{HuangPRR2023}, or variational quantum eigensolvers \cite{peruzzo2014variational}, become crucial. In particular, quantum control-inspired algorithms have demonstrated themselves as invaluable approaches for addressing complex quantum control problems by leveraging classical optimizers \cite{IvanoPRL2020,SarovarPRR2021,Sun_2022,HuangPRR2023}.

Moreover, facilitating the long-distance transfer of quantum states turns out to be key in the context of large-scale quantum information-processing endeavors. In this context, the utilization of edge states within the SSH model provides a compelling avenue, with these edge states serving as a quantum channel that enables the implementation of steadfast quantum state transfer between distant qubits. The realization of robust edge-state transfer is achievable by adiabatic passage \cite{edge-state-CTAP,edge-state-sshvariant}, drawing inspiration from the original concept of Thouless pumping  \cite{Thouless-pumping}. 
Recent experimental achievements have further confirmed the adiabatic robust transfer of atomic momentum states across synthetic lattices of Bose-Einstein condensed states \cite{edge-state-experiment}. While adiabatic passage exhibits commendable resistance against uncorrelated disorder, its vulnerability to environmental noise remains a challenge. To circumvent this limitation, nonadiabatic transfer protocols \cite{DAngelis_2020, zurita2023fastquantumtransfer} have emerged as effective strategies. Operating within shorter time frames, these protocols mitigate the adverse impact of decoherence. However, it is important to note that Thouless pumping is not generically robust to nonadiabatic effects, as revealed in nonquantized charge-transport scenarios \cite{nonadiabatic-pumping} within a Floquet framework. Consequently, the pursuit of transfer schemes that combine both speed and robustness remains as a promising path in advancing the field of quantum state-transfer methodologies.

In this paper, we present a hybrid analog-digital way to design the nonadiabatic protocol for transferring edge states in an SSH model with odd-number sites. Building on the concept of shortcuts to adiabaticity \cite{STA_2019}, our method involves deriving the counterdiabatic (CD) terms through a variational approach \cite{Sels_2017}, also referred to as quantum transitionless driving \cite{Berry_2009}. This implies the incorporation of approximate two-body and many-body interactions \cite{claeys2019floquet,Xie2022Hubbard}, easily implemented in digitized adiabatic computation \cite{Narendra2021prappl}. 
By incorporating these CD terms, specifically the next-to-nearest neighbor (NNN) hopping terms within sublattice A, into the original Hamiltonian, we achieve an efficient representation of the adiabatic dynamics. This ensures that the state transfer consistently aligns with the zero-energy (instantaneous) eigenstate of the original Hamiltonian.
Furthermore, we leverage DQS within variational quantum circuits to obtain optimal CD driving during the transfer process. Our hybrid analog-digital protocol remains remarkably resilient even when subjected to disorder.

The subsequent sections of this paper are organized as follows. In \seclabel{sec:ssh}, we introduce the SSH model for an odd number of sites and delve into the characteristics of its edge state. \seclabel{sec:analog-way} outlines the nonadiabatic protocol for transferring edge states in the SSH model by using nested commutators (NCs), elucidating CD driving located as NNN hopping terms along sublattice A. In~\seclabel{sec:digital-implementation}, a detailed description of the quantum circuit implementation of the time evolution is provided, where we find the optimal CD driving digitally. In~\seclabel{sec:experimental-implementation}, we discuss the feasibility of various experimental platforms and robustness against disorder. Finally, we conclude in~\seclabel{sec:conclusions}. This work not only provides an efficient protocol for the edge-state transfer that is applicable in different quantum platforms but also opens a door to the synergy of quantum control and DQS.

\section{SSH model: Hamiltonian and Preliminaries}\label{sec:ssh}

The Su-Schrieffer-Heeger (SSH) model, initially proposed in 1979 to depict solitons in polyacetylene~\cite{su1979solitons}, describes a 1D chain formed by two sublattices with chiral symmetry. For the case of an odd number $2N-1$ of sites, comprised of $N$ sites on sublattice A and $N-1$ sites on sublattice B, the Hamiltonian is given by
\begin{equation}\label{eq:ssh}
  H_0(t) = t_2(t)\sum_{j=1}^{N-1}c^\dagger_{2j}c_{2j-1} + t_1(t)\sum_{j=1}^{N-1}c^\dagger_{2j+1}c_{2j} + \hc,
\end{equation}
where $c^\dagger_i$ ($c_i$) creates (annihilates) a fermion in the $i$th site of the chain. Sites with odd (even) indices belong to sublattice A (B). 
The alternative nearest-neighboring hopping terms $t_2(t)$ and $t_1(t)$ represent the intra- and inter-unit-cell interaction, respectively.
This Hamiltonian, with $2N-1$ sites, exhibits chiral symmetry~\cite{asboth2016ashort}, resulting in a symmetric spectrum with $2N-2$ nonzero-energy levels and one zero-energy level. 
In particular, the exact zero-energy eigenstate 
\begin{equation}\label{eq:eigenstate-zeroenergy-form}
  \ket{\Phi_0(t)} = \frac{1}{\mathcal{N}}\left[\sum_{i=1}^N a_i \ket{A_i} + \sum_{i=1}^{N-1} b_i \ket{B_i}\right],
\end{equation}
has components with the site-occupation amplitudes
 \begin{equation}\label{eq:eigenstate-zeroenergy}
 \begin{aligned}
b_{i} &= 0, &&\forall i\in[1,N-1],\\
 a_{i} &= \left[-\frac{t_2(t)}{t_1(t)}\right]^{i-1}, &&\forall i\in[1,N],
\end{aligned}
 \end{equation}
where $\mathcal{N}^2=\sum\nolimits_{i=0}^{N-1}[t_2(t)/t_1(t)]^{2i}$
is the normalization factor. Here, Eq.~\eqref{eq:eigenstate-zeroenergy} implies that the zero-energy eigenstate only has nonzero probabilities on sublattice-A sites, which satisfies chiral symmetry.
\begin{figure}[!bt]
 \centering
 \includegraphics[width=\linewidth]{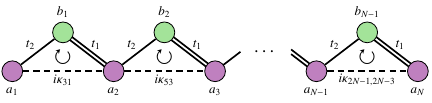}%
\caption{A schematic configuration of an SSH chain. The edge state can be nonadiabatically transferred from the left side to the right one by adding the hopping between the neighboring sites of sublattice A.}\label{fig:ssh}
\end{figure}

Making use of the state $\ket{\Psi(t)}$ evolving along with $\ket{\Phi_0(t)}$, one can adiabatically transfer the state from the left at the initial time $t=0$ to the right bound at the final time $t=T$ (see~\figlabel{fig:ssh})
when the boundary conditions $t_1(0)/t_2(0) \gg 1$, $t_1(T)/t_2(T) \ll 1$, and the general adiabatic condition $T \gg \pi/t_0 $ are satisfied \cite{edge-state-CTAP}. Without loss of generality, we set $t_1(t)=\lambda(t)$ as our control function and $t_2(t)=\Lambda-\lambda(t)$, with a constant $\Lambda$; e.g., applying the hoppings~\cite{DAngelis_2020}
\begin{equation}\label{eq:hopping_cos}
  t_{1,2}(t) = t_0(1\pm\cos\Omega t),
\end{equation}
where $\Omega=0.01t_0$, $\Lambda=2t_0$, $\dot{\lambda}(t)=-\Omega t_0\sin\Omega t$ and, for simplicity, we set $\lambda \equiv \lambda(t)$ in the text. Defining the transfer fidelity
\begin{equation}\label{eq:F}
 F(t)=|\braket{\Phi_0(t)|\Psi(t)}|^2,
\end{equation}
which is used to characterize the extent to which the state evolves along the instantaneous eigenstate in the interval $t\in [0, T]$, we can obtain the fidelity robustness $F(T) > 99.9\%$ at $T \equiv \pi / \Omega = 100\pi /t_0$, fulfilling the adiabatic condition mentioned above.

Indeed, it is worth noting that previous efforts have primarily focused on expediting the slow adiabatic protocol for edge-state transfer by using inverse-engineering methods \cite{DAngelis_2020,Ban_2018}. While inverse engineering provides the advantage of flexibility in designing the time-dependent couplings $t_{1,2}(t)$ (if it is allowed experimentally), it does run the risk of exciting other intermediate states, rather than the zero-energy eigenstate. In contrast, CD driving offers the distinct advantage of ensuring adiabatic following of the original Hamiltonian \cite{Berry_2009,Sels_2017,claeys2019floquet,Narendra2021prappl} but within a shorter time frame. In what follows, we shall focus on the optimal CD terms, providing a promising pathway to achieve more robust and high-fidelity state transfers, even in the presence of noise and other experimental imperfections.

\section{Nonadiabatic edge-state transfer} \label{sec:analog-way}

\subsection{Approximate CD contribution}\label{sec:ssh-analytical}

In this section, we provide a speed-up protocol inspired by CD driving where the state evolution is always along $\ket{\Phi_0(t)}$, by canceling the nonadiabatic transitions from all instantaneous eigenstates. This leads to the total Hamiltonian $H(t)= H_0(t) + H_{\text{cd}}(t)$. However, instead of finding the exact CD terms which requires spectral knowledge, we apply the adiabatic gauge potentials \cite{KOLODRUBETZ20171} to achieve the approximate CD terms
\begin{equation}\label{eq:Hcd}
H^{(l)}_\text{cd}=\dot{\lambda}\mathcal{A}_\lambda^{(l)}
\end{equation}
in the preselected form with the approximate gauge potential expanded in terms of NCs~\cite{Sels_2017, claeys2019floquet} as
\begin{equation}\label{eq:gauge_exp}
\mathcal{A}_\lambda^{(l)}=i\sum_{k=1}^l\alpha^{(l)}_k\underbrace{[H_0,[H_0,\dots,[H_0}_{2k-1},\partial_\lambda H_0]]],
\end{equation}
where $\alpha_k^{(l)}\equiv\alpha_k^{(l)}(t)$, with $k\in\{1,\dots, l\}$ and $l$ being the order of the expansion, are time-dependent coefficients. In the limit of $l\rightarrow\infty$, Eq. \eqref{eq:gauge_exp} becomes the exact gauge potential. Consequently, considering a larger value of $l$ makes the approximate CD terms approach the exact CD driving more closely. The coefficients  $\alpha_k^{(l)}$ are determined by minimizing the action $S_l=\trace{G_l^2}$ with $G_l=\partial_\lambda H_0-i[H_0,\mathcal{A}_\lambda^{(l)}]$ (see details in~\applabel{sec:Appendix2NC}). The first-order NC, 
$\mathcal{A}_\lambda^{(1)}=i\alpha_1^{(1)}[H_0,\partial_\lambda H_0]$, is given by 
\begin{equation}\label{eq:first_order_nested_commutator}
  \mathcal{A}_\lambda^{(1)} =i \kappa^{(1)} \left[ \sum_{j=1}^{N-2}c^\dagger_{2j+2}c_{2j}-\sum_{j=1}^{N-1}c^\dagger_{2j+1}c_{2j-1} \right] + \hc, 
\end{equation}
where $\kappa^{(1)}= -\Lambda\dot{\lambda} \alpha_1^{(1)}$, and $\alpha_1^{(1)}$ is obtained by minimizing $S_1=\trace{G_1^2}$. In this scenario, the CD driving is situated as the NNN hopping between odd-odd and even-even sites with the same absolute strength $|\kappa^{(1)}|$. More detailed analytical solutions on the first-order NC for a general SSH chain with $2N-1$ sites can be found in~\applabel{sec:Appendix2NC}.

As the nonadiabatic transfer protocol needs the NNN hoppings, as shown in Eq.~\eqref{eq:first_order_nested_commutator}, chiral symmetry~\cite{PhysRevB.99.035146, DAngelis_2020} is broken. However, aiming at transferring the edge state along the zero-energy eigenstate [Eq. \eqref{eq:eigenstate-zeroenergy}], the probabilities in the even sites are zero leading to zero hopping between even and even sites. Consequently, the first-order NC terms are simplified to only be located between the neighboring sublattice A.

Subsequently, the second-order NC term goes as
\begin{equation}
\label{eq:second_order_nested_commutator}
\mathcal{A}_\lambda^{(2)} = i \left(\alpha_1^{(2)}[H_0,\partial_\lambda H_0] + \alpha_2^{(2)}[H_0, [H_0, [H_0,\partial_\lambda H_0]]] \right),
\end{equation}
where
\begin{wide}
\begin{equation}\label{eq:nested_commutator_2nd}
\begin{split}
 [H_0,[H_0,[H_0,\partial_\lambda H_0]]]&=\Lambda(\Lambda-\lambda)^2\left[4\sum_{j=1}^{N-2}(c^\dagger_{2j+1}c_{2j-1}-c^\dagger_{2j+2}c_{2j})+c^\dagger_{2N-1}c_{2N-3}\right] \\
&+\Lambda\lambda^2\left[\sum_{j=1}^{N-2}(3c^\dagger_{2j+3}c_{2j+1}+c^\dagger_{2j+1}c_{2j-1}-4c^\dagger_{2j+2}c_{2j})+c^\dagger_{2N-1}c_{2N-3}\right] \\
 &+4\Lambda\lambda(\Lambda-\lambda)\left[\sum_{j=1}^{N-3}(c^\dagger_{2j+3}c_{2j-1}-c^\dagger_{2j+4}c_{2j})+c^\dagger_{2N-1}c_{2N-5}\right]-\hc
\end{split}
\end{equation}
\end{wide}%
The variational parameters $\alpha_{1}^{(2)}$ and $\alpha_{2}^{(2)}$ are obtained by minimizing $S_2=\trace{G_2^2}$ where $G_2$ is both $\alpha_{1}^{(2)}$ and $\alpha_{2}^{(2)}$ dependent. As a result, the CD driving resulting from the second-order NC $H_\text{cd}^{(2)}(t) = \dot{\lambda}\mathcal{A}_\lambda^{(2)}$ involves fourth-nearest-neighboring terms among sublattices A and B sites. Similar to the analysis on the first-order NC, the hoppings between the NNN and the fourth-nearest-neighboring terms on sublattice B can be neglected for the second-order NC due to the transfer trajectory along the zero-energy eigenstate [Eq. \eqref{eq:eigenstate-zeroenergy}].

Indeed, obtaining higher orders of NC terms requires significantly more effort both in terms of the computational resources needed as well as in terms of physical implementation. The gauge-potential expansion $\mathcal{A}_\lambda^{(l)}$ requires calculating sets of NCs as well as $l$ time-dependent coefficients $\alpha^{(l)}_k$, to be computed after action-$S_l$ minimization. After that, the time evolution for the edge-state transfer can be solved as a system of coupled differential equations with several embedded time dependencies, which turns out to be challenging for larger systems.

On top of that, as detailed in~\applabel{sec:d-NC-2N-1}, the NC formalism for the SSH case introduces new hoppings starting from the NNNs to the $2l$th neighbours. Therefore, computing this quantity for higher orders involves more interactions constrained in the same sublattice. In general, for a system with $2N-1$ sites, at least $\mathcal{A}_\lambda^{(N-1)}$ should be computed to account for all possible interactions. From $\mathcal{A}_\lambda^{(N)}$, higher orders will start accumulating new contributions to the hoppings already considered with the lower-order NC. Consequently, all possible long-range hoppings in the same sublattices are needed to accomplish the exact gauge potential, which is challenging and impractical for realistic setups.

Given these challenges, the exploration of simpler strategies or approximations may be necessary for practical implementation. It is common practice to strike a balance between accuracy and computational complexity in quantum simulations, particularly when dealing with higher-order terms.

\subsection{Numerical results with simplified CD driving}
\label{sec:ssh-numerical}

\begin{figure*}[!tb]
 \centering
 \includegraphics{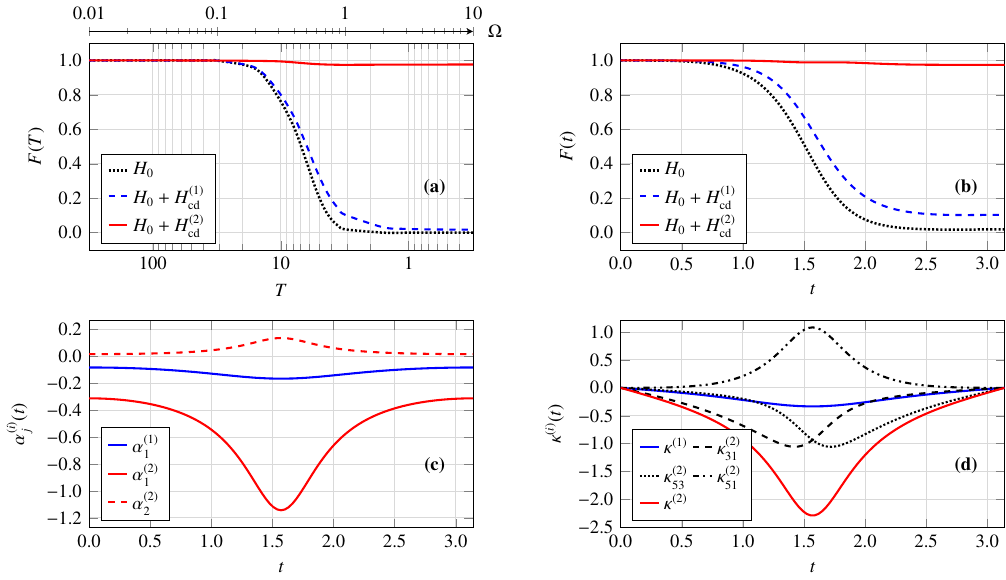}%
\caption{The use of different schemes with the Hamiltonians $H_0$, $H_0 + H_\text{cd}^{(1)}$, and $H_0 + H_\text{cd}^{(2)}$ to transfer the edge state of an SSH chain with five sites. \textbf{(a)} The fidelity robustness $F(T)$ in the dependence of the operation time $T$ ranging from the adiabatic ($\Omega=0.01t_0$) to the nonadiabatic regime (up to $\Omega=10t_0$). \textbf{(b)} The transfer fidelity $F(t)$ during the transfer, where $T=\pi$ and $\Omega=t_0$. \textbf{(c)} The variational parameters $\alpha^{(1)}_1$, $\alpha^{(2)}_1$, and $\alpha^{(2)}_2$ derived by minimizing the action $S_l=\trace{G_l^2}$ ($l=1,2$). \textbf{(d)} The sublattice-A hoppings for the edge-state transfer: different strengths $\kappa^{(2)}_{31}$, $\kappa^{(2)}_{53}$, and $\kappa^{(2)}_{51}$ [Eq.~\eqref{eq:Hcd_5sites_simplified}] and equal strengths $\kappa^{(2)}=-\Lambda\dot{\lambda}\alpha^{(2)}_1$ [Eq. (\ref{eq:equal_kappa_5sites})] from the second-order NC in comparison with $\kappa^{(1)}=-\Lambda\dot{\lambda}\alpha^{(1)}_1$ derived from the first-order NC and $\Lambda=2t_0$ with $t_0=1$.}
\label{fig:5sites}
\end{figure*}

Next, we exemplify our protocol using a simple chain with five sites (denoted by $N=3$). The CD driving terms derived from the first-order and the second-order NC terms are obtained as
\begin{align}
\label{eq:Hcd_5sites}
H^{(1)}_\text{cd} &= i  \Lambda \dot{\lambda} \alpha_1^{(1)} (c_3^\dagger c_1 + c_5^\dagger c_3 - c_4^\dagger c_2) + \hc
 \\
\begin{split} \label{eq:Hcd_5sites_2}
H^{(2)}_\text{cd} &= i  \Lambda \dot{\lambda} [\alpha_1^{(2)} + \alpha_2^{(2)}(t_1^2 + 4 t_2^2)] c_3^\dagger c_1
\\ & + i  \Lambda \dot{\lambda} [\alpha_1^{(2)} + \alpha_2^{(2)}(4t_1^2 + t_2^2)] c_5^\dagger c_3
\\ & - i  \Lambda \dot{\lambda} [\alpha_1^{(2)} + 4\alpha_2^{(2)}(t_1^2 + t_2^2)] c_4^\dagger c_2
\\ &+ 4i \Lambda\dot{\lambda} \alpha_2^{(2)} t_1 t_2 c_5^\dagger c_1 + \hc,
\end{split}
\end{align}
where the second-order NC introduces the NNN hoppings, resulting from the competition between $\alpha_1^{(2)}$ and $\alpha_2^{(2)}$ and the emergence of fourth-nearest-neighboring hopping.
By employing the hopping terms in the form of Eq. \eqref{eq:hopping_cos}, with $\Omega = 0.01 t_0$,  the adiabatic transfer for this five-site chain is achieved at $T= \pi/\Omega = 100\pi$, with $t_0=1$ throughout. However, shortening the operation time into the nonadiabatic regime induces the unwanted excitation. In~\figlabel{fig:5sites}(a), transfer protocols for five sites, including the use of $H_0$, $H_0+H_\text{cd}^{(1)}$, and $H_0+H_\text{cd}^{(2)}$, present the fidelity robustness $F(T)$ as a function of the operation time $T$ (equivalently, in $\Omega$). Adding $H_\text{cd}^{(2)}$ to $H_0$ exhibits high-fidelity transfer even in the nonadiabatic regime. In particular, we choose the specific case of $\Omega = t_0$, so that $T=\pi$, where the adiabatic condition is not satisfied.
As shown in~\figlabel{fig:5sites}(b), the protocol with $H_0 + H_{\text{cd}}^{(2)}$ retains high fidelity during the transfer. \figlabel{fig:5sites}(c) shows the variational parameters $\alpha_1^{(1)}$ and $\alpha_{1,2}^{(2)}$, the derivation of which can be found in \applabel{sec:Appendix2NC}.
The CD driving from the second-order NC varies in different locations, manifested as the competition of terms of $\alpha_1^{(2)}$ and  $\alpha_2^{(2)}$, as indicated in Eq. \eqref{eq:Hcd_5sites}.

The CD driving from the second-order NC can be further simplified with the following two procedures:
(i) neglecting NNN hopping terms on sublattice B due to the transfer trajectory along the zero-energy state; and (ii) performing CD driving only for NNN hopping terms within sublattice A while neglecting longer-range hopping terms. As a first step, the hoppings between two and four sites (couplings on sublattice B) in Eq.~\eqref{eq:Hcd_5sites_2} are neglected, resulting in a simplified form, 
\begin{equation}\label{eq:Hcd_5sites_simplified}
\tilde{H}^{(2)}_\text{cd} = -i \kappa^{(2)}_{31} c_3^\dagger c_1 - i \kappa^{(2)}_{53} c_5^\dagger c_3 - i\kappa^{(2)}_{51}c_5^\dagger c_1 + \hc
\end{equation}\color{black}
with the symmetric CD driving strengths being $\kappa^{(2)}_{31} = -\Lambda \dot{\lambda} [\alpha_1^{(2)} + \alpha_2^{(2)}(t_1^2 + 4 t_2^2)]$, $\kappa^{(2)}_{53} = -\Lambda \dot{\lambda} [\alpha_1^{(2)} + \alpha_2^{(2)}(4t_1^2 +  t_2^2)]$, and $\kappa^{(2)}_{51} = -4\Lambda \dot{\lambda}\alpha_2^{(2)}t_1t_2$.
The simplified scheme $\tilde{H}^{(2)}_\text{cd}$ [Eq. \eqref{eq:Hcd_5sites_simplified}], with the CD driving shown in~\figlabel{fig:5sites}(d), yields the exact same transfer fidelity as 
the original $H^{(2)}_\text{cd}$, which includes all terms [Eq. \eqref{eq:Hcd_5sites}],  shown in~\figlabel{fig:5sites}(b) (solid red).
Second, continuing to simplify CD driving strengths by neglecting $\alpha_2^{(2)}$ terms leads to
\begin{equation}
\label{eq:equal_kappa_5sites}
\kappa^{(2)} =\kappa'^{(2)}_{31} = \kappa'^{(2)}_{53} = -\Lambda \dot{\lambda} \alpha_1^{(2)},
\end{equation}
as shown in~\figlabel{fig:5sites}(d). Equal CD driving in different locations can also fulfill the transfer in a five-site chain with high fidelity very similar to (indistinguishable from)~\figlabel{fig:5sites}(b) (solid red).
This indicates that for a five-site chain, applying the hopping terms only between neighboring sublattice A from the second-order NC as the CD driving maintains the transfer with high fidelity, although the chiral symmetry of the SSH chain is broken \cite{PhysRevB.99.035146,DAngelis_2020}.

For longer chains, the contribution from higher-order NC is necessary. The exact CD driving $H_\text{cd}^{(l)}(t) = \dot{\lambda}\mathcal{A}_\lambda^{(l)}$ should be derived rigorously from $\mathcal{A}_{\lambda}^{(l)}$ with the limit $l\rightarrow\infty$ to retrieve the actual gauge potential. However, to this end, one needs to derive $\alpha_1^{(l)}, \alpha_2^{(l)},\dots,\alpha_l^{(l)}$ with its corresponding large computational cost. To further simplify the protocol for a $2N-1$ chain, we aim to find 
\begin{equation}
\label{eq:Hcd_5sites_infinity}
H_\text{cd} = -i \sum_{j=1}^{N-1} \kappa_{2j+1, 2j-1} c^\dagger_{2j+1} c_{2j-1} + \hc,
\end{equation}
and seek for optimal ways to uncover $\kappa_{2j+1, 2j-1}$ so that CD driving is only attributed to NNN hoppings in sublattice A.

\section{Digital simulation and Optimization}
\label{sec:digital-implementation}

Given the complexity of calculating high-order NC terms, particularly the variational parameter $\alpha^{(l)}_1$, DQS emerges as an efficient way 
to learn the prefactor for the NC, i.e., the CD driving.
In~\seclabel{sec:2N-1sites-simulator}, a general scheme for digital simulation of a $2N-1$ SSH chain in quantum circuits is listed, where the general Hamiltonian is digitally encoded and digitized. Aiming to achieve high-fidelity transfer, we separately test two different cost functions by making use of the transfer fidelity and a Hellinger distance-based method, deriving the optimal CD driving concurrently. 
In~\seclabel{sec:results}, we exemplify the edge-state transfer in ($5$-$15$)-site chains by using \textsc{qiskit}~\cite{qiskit} to simulate ideal quantum circuits with \textsc{ibmq\_qasm\_simulator}~\cite{ibm}.

\subsection{Encoding and classical optimizers} \label{sec:2N-1sites-simulator}

The edge-state transfer in an SSH chain of $2N-1$ sites requires at least $n=\lceil\log_2 (2N-1)\rceil$ qubits to encode the process by using the Hamiltonian $H_0$ [Eq. \eqref{eq:ssh}]. Such a Hamiltonian has to be padded as
\begin{equation}\label{eq:h_pad}
 H_0^{(c)}(t) =
 \begin{bmatrix}
 H_0(t) & 0 \\
 0     & 0 \\
 \end{bmatrix},
\end{equation}
with $H_0^{(c)}(t)\in\mathcal{M}_{2^n}(\mathbb{R})$.

Our goal is to transfer the edge state nonadiabatically by introducing CD driving, represented by NNN interactions acting exclusively on sublattice A
\begin{equation}
\label{eq:hcd_circ_no24}
\begin{aligned}
  H_{\text{cd}}^{(c)} = -i \sum_{j=1}^{N-1} \kappa^{(c)}_{2j+1, 2j-1}c_{2j+1}^\dagger c_{2j-1} + \hc
\end{aligned}
\end{equation}
Here, the superscript $(c)$ denotes that the optimal but unknown prefactor $\kappa^{(c)}$ is derived from the circuit, approaching $\kappa^{(l)}$ $(l \rightarrow \infty)$. The subscript indicates the location of the hopping term. 
It is worth noting that the form of this driving is based on the first-order NC as shown in Eq. \eqref{eq:first_order_nested_commutator}.
To digitally implement the time evolution, both Hamiltonians must be decomposed into the Pauli basis. The decomposition can be performed using methods presented in previous works, such as Refs.~\cite{romero2023paulicomposer,HuangPRR2023,Liu2022digital}. In~\applabel{sec:pauli_decomposition}, general decompositions of Hamiltonians are included, considering all possible NNN interactions between sublattice A sites.

In the circuit, the state is evolved as $\ket{\Psi(t)}=U(t, 0)\ket{\Psi(0)}$ with $t\in [0, T]$. The initial state is $\ket{\Psi(0)} = \ket{\Phi_0(0)}$ (the probability is $1$ at the first site and $0$ otherwise), which can be encoded in the computational basis as $\ket{\Psi(0)}=\ket{0\dots0}=\ket{0}$. We employ product formulas for exponentials of commutators \cite{chen2022efficient}, digitally exploring them for 1D fermion chains with nearest- and next-nearest-neighbor hopping terms. In our work, the circuit can be constructed with Pauli gates to digitize the time evolution of Hamiltonian $H_0^{(c)} + H_{\text{cd}}^{(c)}$ with the unitary operator
\begin{equation}\label{eq:trotter_circuit}
  U(T,0)\approx\prod_{k=1}^r\exp\left[{-iH_0^{(c)}(k\Delta t)\Delta t}\right]\exp\left[{-iH_{\text{cd}}^{(c)}(k\Delta t)\Delta t}\right].
\end{equation}
This is achieved by computing the Trotter-Suzuki decomposition~\cite{trotter1959product,suzuki2005finding} up to the first order, with $r\coloneqq T/\Delta t$ steps, where $\Delta t$ is the interval of each Trotter step.
We discretize $\kappa^{(c)}_{2j+1,2j-1}$ as a set of $r+1$ parameters,
\begin{equation}\label{eq:params}
 \vec{\kappa}^{(c)}_{2j+1,2j-1}(t)\coloneqq[\kappa^{(c)}_{2j+1,2j-1}(0), \kappa^{(c)}_{2j+1,2j-1}(\Delta t), \dots, \kappa^{(c)}_{2j+1,2j-1}(T)], ~~
\end{equation}
with boundary conditions $\kappa^{(c)}_{2j+1,2j-1}(0)=\kappa^{(c)}_{2j+1,2j-1}(T)=0$, so that $r-1$ parameters will be optimized. \color{black}
The digital form of the Hamiltonian $H_0^{(c)}+H_{\text{cd}}^{(c)}$, i.e., Eq.~\eqref{eq:trotter_circuit}, is implemented in the quantum circuit as depicted in~\figlabel{fig:qc}. Our goal is to generate the target state $\ket{\Psi(T)}$ as close as possible to $\ket{\Phi_0(T)}=\ket{2N-2}$ (the probability is $1$ at site $2N-1$ and $0$ otherwise), up to the set of optimal parameters $\vec\kappa^{(c)}$.
\begin{figure}[!tb]
 \centering
 \includegraphics[width=.95\linewidth]{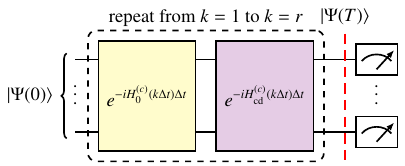}%
\caption{The circuit implementation of the state evolution by using the unitary operator $U(T, 0)$~[Eq.\eqref{eq:trotter_circuit}] Trotterizing the Hamiltonian which is the sum of $H_0^{(c)}$ (yellow) and $H_{\text{cd}}^{(c)}$ (purple).
}\label{fig:qc}
\end{figure}%

In general, various classical optimizers, such as the adaptive moment estimation algorithm (Adam), and constrained optimization by linear approximations (COBYLA), are available in the selection pool, with their pros and cons. Simultaneous perturbation stochastic approximation (SPSA), a gradient-descent method for optimizing systems with multiple unknown parameters, is a suitable simulation-optimization method with a relatively small number of measurements of the objective function, appropriate to be applied in the SSH model in this context. In order to ease the optimization routine utilizing \emph{warm starting}, we initialize the parameters at each time step with an initial guess proportional to $\kappa^{(1)}$. Additionally, adaptive bounds are set for the $r-1$ parameters to optimize CD driving.

To guarantee the state evolution along the zero-energy eigenstate, we apply two cost functions respectively: one designed from the transfer fidelity and the other retrieved from the measurement in quantum circuits. The first cost function, $C(\vec{\kappa}^{(c)}(t))=F(T)$ [Eq. \eqref{eq:F}], where $\vec{\kappa}^{(c)}$ collects all the NNN couplings $\kappa^{(c)}_{2j+1,2j-1}$ $\forall j\in[1,N-1]$, conceived from the control problem itself, maximizes the fidelity at the final time $T$ between the zero-energy eigenstate and the state derived in the circuit. 
These overlaps can be computed in quantum circuits due to SWAP tests~\cite{garcia2013swap,cincio2018learning},where the required number of qubits equals the number needed to encode both overlapping states~\cite{knill2007optimal,havlicek2019supervised}. 
To use a reduced number of qubits and to get optimization from utilizing a finite number of measurements in quantum circuits, we apply the Hellinger distance~\cite{hellinger1909neue} (also called the Jeffreys distance~\cite{jeffreys1946invariant}) which quantifies the similarity between two given probability distributions, $P$ and $Q$, over a measure space $\mathcal{X}$, with corresponding probability density functions $p(x)$ and $q(x)$, where $x\in\mathcal{X}$. Its integral form is defined as
\begin{equation}\label{eq:int_hell}
 H^2(P,Q) =1-\int_\mathcal{X}\sqrt{p(x)q(x)}\text{d}x,
\end{equation}
where $0\leq H^2(P,Q)\leq 1$ using the Cauchy-Schwarz inequality. For discrete distributions of $m$ values, let $P=\{p_i\}_{i=0}^{m-1}$ and $Q=\{q_i\}_{i=0}^{m-1}$, so that Eq.~\eqref{eq:int_hell} becomes
\begin{equation}\label{eq:discrete_hell}
 1-H^2(P,Q)=\sum_{i=0}^{m-1}\sqrt{p_iq_i}.
\end{equation}
Among all the $2^n$ possible outcomes in the circuit (ranging from $\ket{0}$ to $\ket{2^n-1}$), we expect to maximize the probability of measuring $\ket{2N-2}$ as the final circuit state. The Hellinger distance $H^2(P(\vec{\kappa}^{(c)}(t)),Q)$ is used as the cost function, where $P(\vec{\kappa}^{(c)}(t))$ is the probability distribution after $N_\text{shot}$ measurements. The target probability distribution $Q$ is defined as $q_{2N-2}=1$ and $0$ otherwise. To get $P(\vec{\kappa}^{(c)}(t))$ as close as possible to $Q$, i.e., to maximize the probability of transferring the edge state, we define the cost function as
\begin{equation}\label{eq:hell_cost}
 \min_{\vec{\kappa}^{(c)}} H^2(P(\vec{\kappa}^{(c)}), Q)\equiv\max_{\vec{\kappa}^{(c)}} p_{2N-2},
\end{equation}\color{black}
which acts analogously to the transfer fidelity [Eq.~\eqref{eq:F}] in the large-$N_\text{shot}$ limit. Therefore, as seen from Eq.~\eqref{eq:hell_cost}, minimizing the Hellinger distance is translated into maximizing the population in the site $a_N$ at $t=T$, which is indeed the main goal of the optimization process.

\subsection{Numerical results and analysis}\label{sec:results}

In this subsection, we give a detailed example of the transfer over SSH chains with different numbers of sites, going from five ($N=3$) to $15$ sites ($N=8$), thus requiring at most $n=4$ qubits for the digitalization in a quantum circuit approaching $\ket{\Phi_0(T)}=\ket{2N-2}$ starting from $\ket{\Phi_0(0)}=\ket{0}$. Hereinafter, we shall work in the nonadiabatic regime $t_0=\Omega=1$ and thus we set $T=\pi$. The Pauli-basis decomposition of Eqs. \eqref{eq:h_pad} and \eqref{eq:hcd_circ_no24} can be computed following~\applabel{sec:pauli_decomposition}. To exemplify the technique, the decomposition for a five-site chain becomes
\begin{wide}
 \begin{align}
 \label{H0c}
  H_0^{(c)}(t) &= \frac{\Lambda-\lambda(t)}{2}(I+Z)IX + \frac{\lambda(t)}{4}[(I+Z)(XX +  YY) + X(XX - YY) + Y(XY + YX)], \\
  H_{\text{cd}}^{(c)}(t) &=\frac{1}{4} [\kappa^{(c)}_{31}(t)(I+Z)Y(I+Z) + \kappa^{(c)}_{53}(t)(YX-XY)(I+Z)].
  \label{Hcdc}
 \end{align}
\end{wide}%
Regarding the NNN hoppings $\vec{\kappa}^{(c)}$, two approaches will be tested: time-symmetric hoppings due to symmetry and equal NNN hoppings inspired by the 1NC matrix form. For the first of these, the NNN hoppings obey $\kappa^{(c)}_{2i+1, 2i-1}(t)=\kappa^{(c)}_{2(N-i)+1,2(N-i)-1}(T-t)$ with $i\in[1,\lfloor N/2\rfloor]$ so that $\lfloor N/2\rfloor(r-1)$ parameters are required. For the latter case, we factor out the hopping strengths of Eq. \eqref{eq:hcd_circ_no24}; thus the NNN hoppings become site independent, i.e., $\kappa^{(c)}_{2j+1,2j-1}(t)\mapsto\kappa^{(c)}(t)$. Consequently, $r-1$ parameters need to be optimized. The latter simplified protocol is easier  to implement.

We Trotterize the time-evolution operator [Eq. \eqref{eq:trotter_circuit}] into $r=22$ steps and run the variational quantum circuit with $1000$, $2000$, $3000$, $5000$, $6000$, and $10000$ optimizer iterations for the chains with $5-15$ sites, respectively. Such a realization is repeated $10$ times to observe the variance (and thus the stability) of our method. To guarantee $\mathcal{O}(\norm{H_0})\sim\mathcal{O}(\norm{H_\text{cd}})$ for feasibility purposes---i.e., the CD driving is in the same order of $t_0$---we take the value of $\kappa^{(1)}$ times a constant $\sigma_\kappa$ as the initial guess, while the optimal parameters are set to lie between a constant $\kappa_\text{bound}$ in the same order of magnitude of the initial guess minima and zero.
For our study, we take $\sigma_\kappa$ a random value within $[4,5]$, $[6,7]$, $[7,8]$, $[9,10]$, $[10,11]$ and $[11,13]$ as well as $\kappa_\text{bound}\in[-2.5, -3, -3.5, -4, -4.5, -5]$ for $5-15$ sites, respectively, to show the stability of the results.

\begin{figure*}[!tbp]
 \centering
 \subfloat{%
 \includegraphics{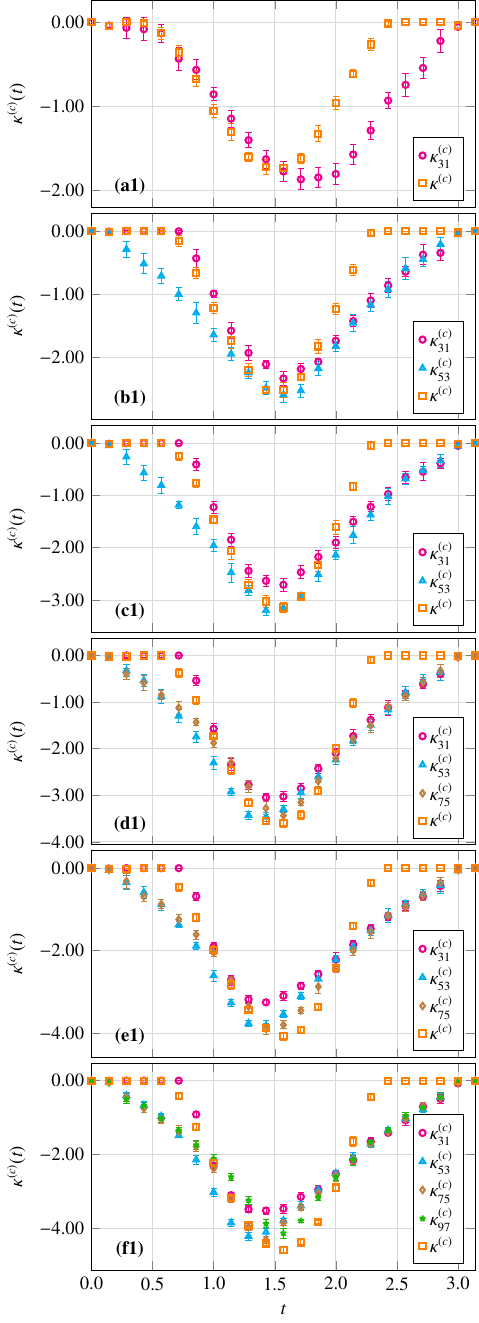}
 }\hspace{1cm}%
 \subfloat{%
 \includegraphics{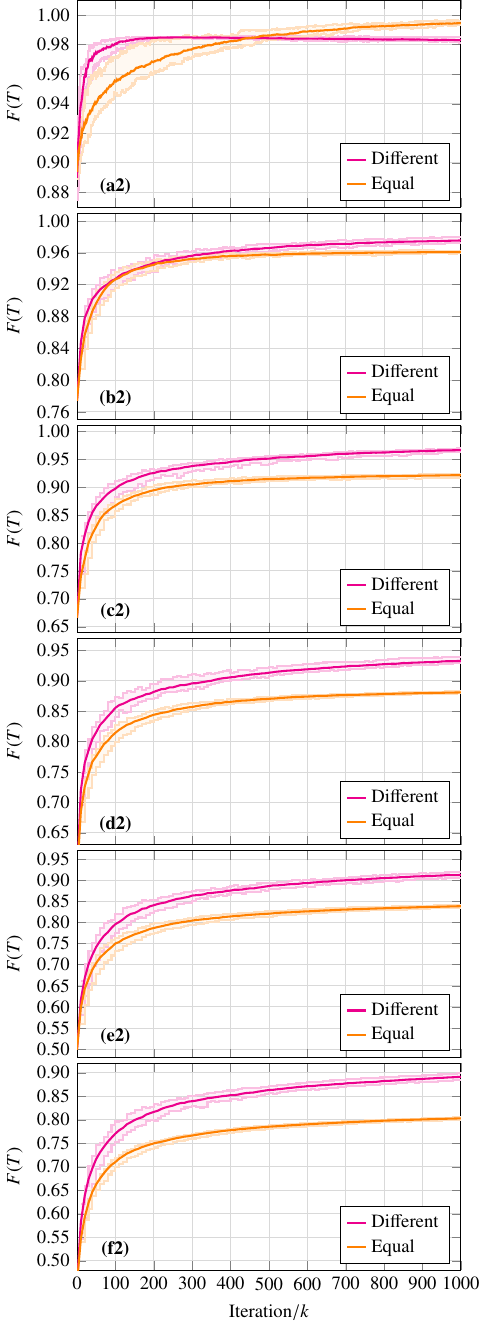}}%
 \caption{From top to bottom: results for chains of {five, seven, nine, $11$, $13$ and $15$ sites} [\textbf{(a)}-\textbf{(f)}, respectively] using both equal (orange squares) and different NNN hoppings (rest) on sublattice A after ten runs. Left: the NNN couplings for each chain are represented in the case of $\Omega=t_0=1$ with the labeled average values and their corresponding standard deviation. Right: the evolution of the cost function $F(t)$ for the chains with five, seven, nine, $11$, $13$ and $15$ sites against the rescaled number of iterations with $k\in[1,2,3,5,6,10]$ in each case, respectively. The shaded areas show the region of the returned minimum and maximum values.}\label{fig:kappa_circ}
\end{figure*}
\begin{figure*}[!t]
    \centering
    \includegraphics{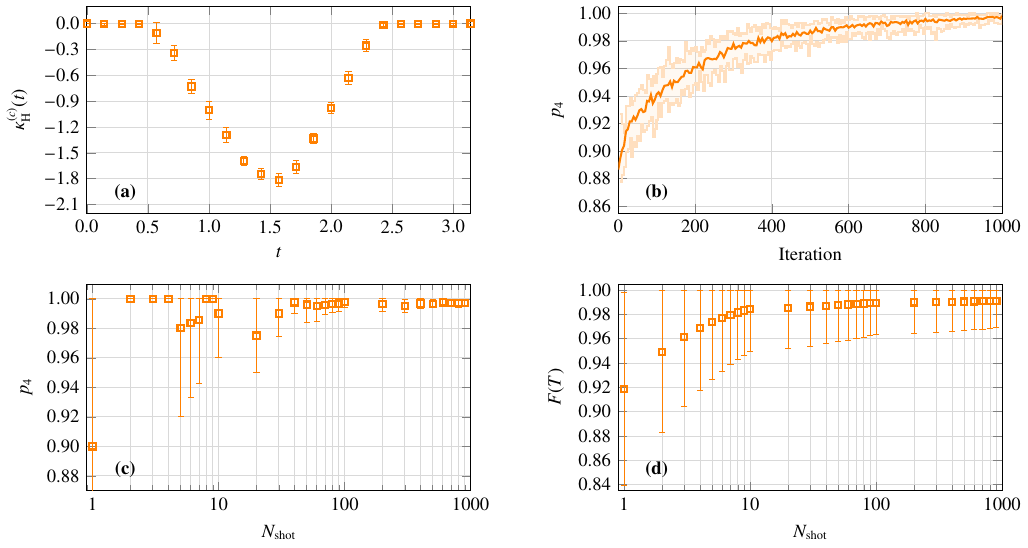}%
    \caption{Using the Hellinger distance as the cost function [Eq.~\eqref{eq:hell_cost}], we find optimal CD driving for a five-site chain from running quantum circuits $10$ times, where $\Omega=t_0=1$, when considering equal NNN hopping terms on sublattice A. \textbf{(a)} $\kappa^{(c)}_\text{H}$. \textbf{(b)}  $p_4$ in terms of the iterations. \textbf{(c)} $p_4$ as a function of the number of measurements $N_\text{shot}$. \textbf{(d)} We simultaneously check $F(T)$ in the presence of the corresponding $N_\text{shot}$ by introducing the optimal driving obtained from the Hellinger distance. Each point derived at a certain $N_\text{shot}$ in (c) and (d) is from running a $1000$-times iteration in the circuit. In (a), (c), and (d), the labeled average values and standard deviation are shown, while the areas of the minimum and maximum returned values are plotted in (b).}\label{fig:kappa_circ_hellinger}
\end{figure*}

Two distinct cost functions are used here. The first is to use the transfer fidelity as a cost function [Eq. \eqref{eq:F}]. The optimal ${\kappa}^{(c)}(t)$ are found in~\figlabel{fig:kappa_circ}(a1-f1), while the evolution of the final fidelity $F(T)$ with respect to the number of iterations considering only NNN hoppings in sublattice A is shown in~\figlabel{fig:kappa_circ}(a2-f2).
The use of only NNN hoppings as CD driving can transfer the edge state with $\mathcal{O}(\norm{H_0})\sim\mathcal{O}(\norm{H_\text{cd}})$ with fidelity $F(T)\sim 90\%$ in a $15$-site chain. If we relax the $\mathcal{O}(\norm{H_0})\sim\mathcal{O}(\norm{H_\text{cd}})$ constraint by increasing the $\sigma_\kappa$ and $|\kappa_\text{bound}|$ values, stronger NNN hoppings will be found. In this way, it is possible to accomplish higher fidelities but at the cost of deriving much stronger NNN hoppings in contrast to $t_{1,2}$. It is also possible to replicate CD driving similar to those derived from inverse engineering \cite{DAngelis_2020} if one uses loose constraints for the strength in the circuits. However, the CD driving strengths have to be limited, depending on the physical setup, in order to avoid strong heating.

The other separate study uses the same procedure and hyperparameters in the quantum circuit but applies the cost function given in Eq.~\eqref{eq:hell_cost} from the Hellinger distance with $N_\text{shot}=1024$. Resembling the study of~\figlabel{fig:kappa_circ}, we obtain similar results shown in~\figlabel{fig:kappa_circ_hellinger}(a-b), which proves the feasibility of using measurement-based quantum circuits.
To estimate the number of measurements needed to acquire good and stable performance of the edge-state transfer, in~\figlabel{fig:kappa_circ_hellinger}(c-d) we illustrate the relation between $p_4$ and the number of measurements $N_\text{shot}$, where each $p_4$ value is obtained by $1000$-times iteration in the circuit. Meanwhile, we also check the transfer fidelity $F(T)$ in the presence of the corresponding $N_\text{shot}$ by using the optimal driving obtained from the Hellinger distance. Running the variational circuit $10$ times, one can obtain an average fidelity above $90\%$ from $N_\text{shot}=1$, and a stable average fidelity above $99\%$ with $N_\text{shot} > 200$.

\section{Discussion} \label{sec:experimental-implementation}

\subsection{Physical feasibility}
Experimental demonstrations of the SSH model have been reported in various platforms, such as quantum dot arrays~\cite{SSH-QDs}, Rydberg atoms~\cite{meier2016observation}, and superconducting circuits~\cite{cai2019observation}, among others.
Recently, dynamically modulated SSH models have shown adiabatic edge-to-edge transport of atomic momentum states in synthetic lattices of Bose-Einstein condensates \cite{edge-state-experiment} on the time-scale of hundreds of microseconds. 
In contrast, our proposed protocol offers robust transfer on a nanosecond timescale before decoherence sets in. Moreover, DQS provides discretized solutions, which are expected to be more easily implemented in experiments.

In the realm of superconducting qubits \cite{mei2018robust}, the engineered hopping strengths, depicted as $t_{1,2}(t)=g_0\mp g_1\cos\Omega t$ similarly to our model in Eq. \eqref{eq:hopping_cos}, provide a promising avenue. Experimental parameters such as $g_1/2\pi=\SI{250}{MHz}$, $g_0 = g_1$, and $T=\pi/\Omega=\SI{0.2}{\micro s}$ in the adiabatic regime ($\Omega=0.01g_1$) have been considered. Notably, shortening the operation time to $T=\SI{2}{ns}$ ($\Omega = g_1$) while keeping $g_1$ constant necessitates the addition of CD hoppings between neighboring sites on sublattice A with a magnitude comparable to $g_1$: these additional hoppings can be implemented by producing a gauge potential via applying geometric phases \cite{Zhang2017-phase}, by means of artificial magnetic fields \cite{magnetic-field-phase}, or by means of angular-momentum states \cite{momentum-phase}, as shown in Fig. \ref{fig:ssh}. The tuning of this CD driving can be achieved smoothly by varying the current passing through the coupler linking corresponding Xmon qubits \cite{qubit-qubit-interaction1, qubit-qubit-interaction2}, where imaginary couplings are indeed realizable in superconducting-qubit setups \cite{qubit-qubit-interaction3}. Experimental validation of this CD driving has been performed by generating gauge-invariant phases in three-level superconducting quantum circuits \cite{Yuexperiment2021,Paraoanuxperiment2021}.
On the other hand, the configuration involving sites $A_i$ and $B_i$ in each unit cell, along with $A_{i+1}$---the neighboring site in sublattice A---constitutes the spatial adiabatic passage block of a three-level system, relevant to systems such as cold atoms trapped in optical lattices \cite{atoms-optical-lattice1,atoms-optical-lattice2,atoms-optical-lattice3} and electrons in quantum dots \cite{CTAP,Ban_2018,Ban_2019}.

\begin{figure*}[!tb]
 \centering
 \includegraphics{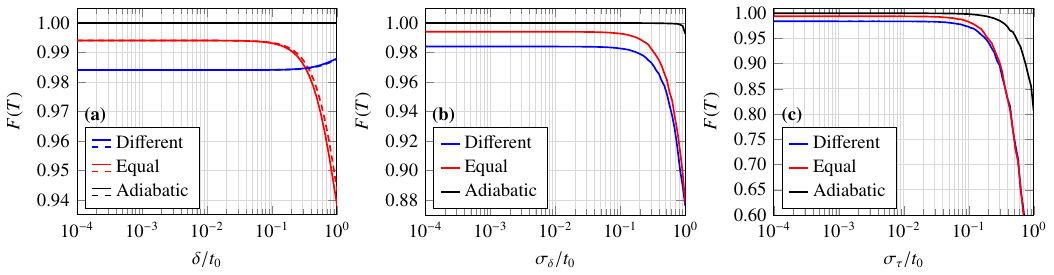}%
 \caption{The robustness of the fidelity $F(T)$ for an SSH chain of five sites using both different and equal NNN hoppings (see~\figlabel{fig:kappa_circ}) as well as in the adiabatic regime $\Omega=0.01t_0$ ($t_0=1$) against \textbf{(a)} the relative strength of the staggered Rice-Mele contribution in Eq.~\eqref{eq:h_diag} in terms of $\delta/t_0$, setting $\delta_j=\tau_j=0$ for $\Delta(t)=\delta$ (solid) and $\Delta(t)=\delta\sin\Omega t$ (dashed), where the fidelities of both parametrizations almost overlap; \textbf{(b)} the diagonal-disorder contribution in Eq.~\eqref{eq:h_diag} in terms of $\sigma_\delta/t_0$, setting $\Delta(t)=\tau_j=0$; and \textbf{(c)} the hopping-disorder contribution of Eq.~\eqref{eq:h_hop} in terms of $\sigma_\tau/t_0$, setting $\Delta(t)=\delta_j=0$. For the latter two [(b) and (c)], the results are averaged over $1000$ runs.}\label{fig:dis}
\end{figure*}

Implementation of the Hamiltonians of Eq. (\ref{H0c}) and Eq. (\ref{Hcdc}) in quantum circuits that include multiple types of three-qubit interactions with precisely calibrated coupling strengths is challenging but promising with continuous progress. For instance, current quantum processing units (QPUs) exhibit connectivities in the hardware of superconducting circuits where each qubit can directly interact with up to three neighbors.
The development of qubit-qubit interactions with all-to-all connections has been carried out in an optical Ising machine hosting adjustable four-body interactions \cite{Kumar2020}. Similar experimental realizations have been conducted in trapped-ion system simulating the dynamics
of four- and six-qubit-interaction Hamiltonians~\cite{Lanyon2011}. We believe that with the development of quantum hardware on the connectivities among qubits, the experimental implementation of digital quantum devices will improve, since the promotion of hardware in near-term devices and quantum control protocols assisted by DQS in respective fields boosts mutual development.

\subsection{Robustness against disorder}

In the context of adiabatic transfer in an SSH model,  external sources such as diagonal disorder acting over on-site energies can break chiral symmetry, while hopping disorder preserves chiral symmetry~\cite{PhysRevB.99.035146}. In our nonadiabatic protocol, chiral symmetry is intentionally broken by adding CD driving between atoms from the same sublattice~\cite{PhysRevB.99.035146,PhysRevLett.123.126401}. To assess the robustness of our nonadiabatic protocol against deviations from the reference Hamiltonian [Eq.~\eqref{eq:ssh}], we examine the probability of transferring the edge state using a modified Hamiltonian: 
\begin{equation}\label{eq:extra_contrib}
H'(t)\coloneqq H_0(t)+H_\text{cd}(t)+H_\text{diag}(t)+H_\text{off-diag},
\end{equation}
where
\begin{equation}\label{eq:h_diag}
 H_\text{diag}(t)=\sum_{j=1}^{2N-1}[(-1)^{j-1}\Delta(t)+\delta_j]c^\dagger_{j}c_j,
\end{equation}
adds a staggered-diagonal potential with $\delta_j\in[-\sigma_\delta,\sigma_\delta]$ denoting a random value encoding a diagonal disorder to the staggered potential contribution. For the Rice-Mele model~\cite{asboth2016ashort}, $\Delta(t)$ can be taken as constant with $\Delta(t)=\delta$ or $\Delta(t)=\delta\sin\Omega t$, where $\delta$ is a real constant~\cite{edge-state-CTAP}. Following similar methods similar to those proposed in Refs.~\cite{mei2018robust,PhysRevB.99.035146}, we consider an additional off-diagonal contribution given by
\begin{equation}\label{eq:h_hop}
 H_\text{off-diag}=\sum_{j=1}^{2N-2}\tau_jc^\dagger_{j+1}c_j+\hc,
\end{equation}
where $\tau_j\in[-\sigma_\tau,\sigma_\tau]$ is a random value encoding a disorder on nearest-neighbor hoppings.

Taking the average CD driving derived from the circuit as shown in~\figlabel{fig:kappa_circ}, we solve the time-dependent Schr\"{o}dinger equation [Eq.~\eqref{eq:extra_contrib}] under the consideration of different scenarios with the aforementioned extra contributions. When a staggered-diagonal term is introduced [\figlabel{fig:dis}(a)], the fidelity degrades more slowly by using different hoppings rather than equal ones, showing that the use of different hoppings provides more robustness for both parametrizations $\Delta(t)=\delta$ and $\Delta(t)=\delta\sin\Omega t$. The staggered Rice-Mele contribution [\figlabel{fig:dis}(a)], the diagonal disorder contribution [\figlabel{fig:dis}(b)], and off-diagonal disorder contribution [\figlabel{fig:dis}(c)] separately present high fidelities over a long range until the ratio of their strengths over $t_0$ is around $0.1$. All of the above results indicate that even though chiral symmetry is broken \cite{PhysRevB.99.035146}, high stability can be preserved in both staggered Rice-Mele contributions and disorder with reasonable strengths.

\subsection{Nonadiabatic transfer solely by CD driving}

\begin{figure*}[!tbp]
 \centering
 \subfloat{%
 \includegraphics{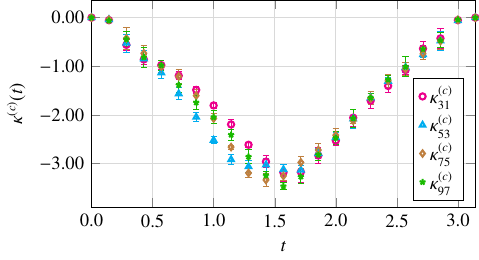}
 }\hspace{1cm}%
 \subfloat{%
 \includegraphics{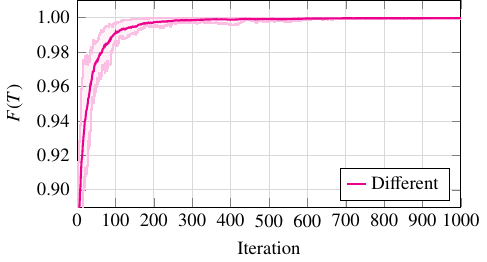}}%
 \caption{Left: the derived different NNN hopping terms on sublattice A for a chain of $15$ sites setting $H_0=0$. NNN couplings are represented in the case of $t_0=1$ and $\Omega=1$ with the labeled average values and their corresponding standard deviation after ten runs. Right: the evolution of the cost function $F(t)$ against the number of iterations, where the shaded areas show the region of the returned minimum and maximum values.}\label{fig:kappa_circ_noh0}
\end{figure*}%
\begin{figure*}[!tb]
 \centering
 \includegraphics{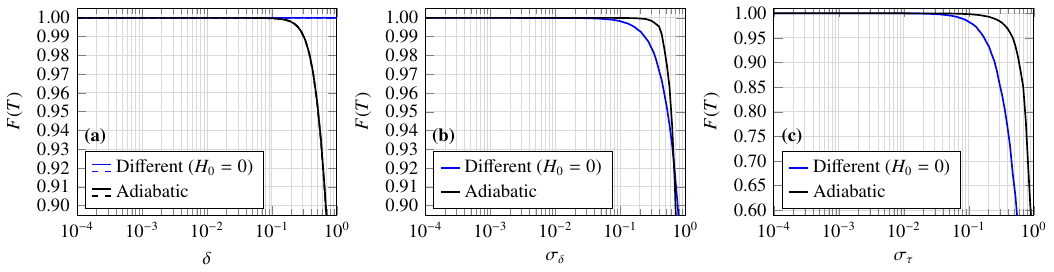}%
 \caption{The robustness of the fidelity $F(T)$ for an SSH chain of $15$ sites using different NNN hoppings on sublattice A, setting $H_0=0$ (blue lines), as well as the expected behavior (black lines) in the adiabatic regime $\Omega=0.01t_0$ ($t_0=1$) against \textbf{(a)} the relative strength of the staggered Rice-Mele contribution in Eq.~\eqref{eq:h_diag} in terms of $\delta$ setting $\delta_j=\tau_j=0$ for $\Delta(t)=\delta$ (solid) and $\Delta(t)=\delta\sin\Omega t$ (dashed), where the fidelities of the two parametrizations almost overlap; \textbf{(b)} the diagonal-disorder contribution in Eq.~\eqref{eq:h_diag} in terms of $\sigma_\delta$, setting $\Delta(t)=\tau_j=0$; and \textbf{(c)} the hopping-disorder contribution of Eq.~\eqref{eq:h_hop} in terms of $\sigma_\tau$ setting $\Delta(t)=\delta_j=0$. For the latter two [(b) and (c)], the results are averaged over $1000$ runs.}\label{fig:dis_noh0}
\end{figure*}

With the intention of achieving a simpler and more reliable setup for edge-state transfer, we propose encoding the entire SSH system information in the CD driving by turning off the reference Hamiltonian ($H_0=0$), thus using the same circuit as shown in~\figlabel{fig:qc} but without the $H_0$ contribution (yellow block). This simplification is motivated by the negligible contribution of NNN hoppings on sublattice B (even-even interactions), satisfying the conditions for transferring the edge state. Moreover, setting $H_0=0$ significantly reduces the depth of the variational quantum circuit needed to extract the desired CD driving, as depicted in \figlabel{fig:qc}. Simultaneously, this simplification eliminates population excitation in sublattice B.

We have conducted experiments with different NNN hoppings, as shown in \figlabel{fig:kappa_circ}, while setting $H_0=0$ for chains with five, seven, nine, $11$, $13$, and $15$ sites. Each case has been tested with 1000 iterations. This simplification has led to a shorter circuit depth, faster convergence, and even better fidelities compared to including $H_0$ in the calculations. The fidelity has consistently exceeded $99\%$.

In \figlabel{fig:kappa_circ_noh0}, we demonstrate the high fidelity of a 15-site chain when $H_0=0$ and different NNN hoppings on sublattice A are used in ten runs. The robustness of this approach against disorder is illustrated in \figlabel{fig:dis_noh0}, where we compare it with the expected behavior in the adiabatic regime ($\Omega=0.01t_0$). Notably, setting $H_0=0$ makes the method stable against Rice-Mele contributions and it exhibits slight performance degradation for the diagonal and hopping disorders but remains stable when the disorder strengths, $\sigma_\delta$ and $\sigma_\tau$, are less than $0.1$.

\section{Conclusions}\label{sec:conclusions}
By synergistically integrating analog and digital strategies, we have presented a nonadiabatic scheme for edge-state transfer in SSH chains with different odd numbers of sites. In particular, we have performed several detailed studies in the nonadiabatic regime $T=\pi$ ($\Omega=t_0$), thus transferring the edge state $100$ times faster than in the regime $T= 100\pi$ ($\Omega=0.01t_0$) set to meet the adiabatic criteria~\cite{DAngelis_2020}. Starting by computing the NCs from approximate gauge potentials, we have found the patterns between the chain length and the location of different orders of NC. Essentially, the CD driving can be interpreted as NNN hoppings among sublattice A sites, although the deviation up to higher order requires a lot of computational cost.
Resorting to DQS, we have employed variational quantum circuits to extract optimal CD driving. The CD driving from the first-order NC formula has been set as the initial values in the circuit with some deviation as degrees of freedom during the optimization process.
For this purpose, two different kinds of cost functions, based on the transfer fidelity and the Hellinger distance, have been used, ultimately yielding high fidelity. The analog analysis to derive the CD driving from NCs displays explicitly the precise location of the additional hoppings required to achieve nonadiabatic driving. Meanwhile, the digital simulation to optimize CD driving facilitates the routines to achieve the additional hoppings. Both methods are mutually reinforcing and complementary, enhancing the ability to effectively navigate nonadiabatic state transfer in the SSH chain.

We have also discussed the practicability of this protocol in different physical platforms such as superconducting qubits and cold atoms trapped in optical lattices---and, especially, the feasibility of generating CD driving utilizing cutting-edge techniques available in these contexts.
In a comprehensive assessment, we have subjected the protocol to rigorous testing, gauging its resilience against an array of supplementary contributions commonly encountered within laboratory settings, including Rice-Mele detuning, diagonal disorder, and hopping disorder. The high fidelity proves the robustness of our method, showing that our protocol is indeed feasible for practical experimental implementation.

\begin{acknowledgments}
We thank Koushik Paul and Pranav Chandarana for the fruitful discussions. 
S.V.R. and Y.B. acknowledge the support of the Basque Government through the Hazitek program (Q4Real project, ZE-2022/00033) and the Gipuzkoako Foru Aldundia–Diputación Foral de Gipuzkoa within the Gipuzkoa Quantum program under the project ``QSIMM: Research in Quantum Simulation for Materials”, 2023-QUAN-000026-01. 
X.C. appreciates the financial support from the Basque Government through Grant No. IT1470-22, through Project Grant No. PID2021-126273NB-I00, funded by Ministerio de Ciencia e Innovacion (MCIN)/AEI/10.13039/5011000 11033, and through the European Regional Development Fund (ERDF) ``A way of making Europe” and the ``ERDF Invest In Your Future,” Nanoscale NMR and complex systems (Grant No. PID2021-126694NB-C21), the European Union (EU) FET Open Grant “Electronic-Photonic Integrated Quantum Simulator Platform” (EPIQUS) (Grant No. 899368), and the Ramón y Cajal program of the Spanish Ministerio de Economía, Comercio y Empresa (MINECO) (Grant No. RYC-2017-22482). 
G.P. was supported by Spain’s MINECO through Grant No. PID2020-117787GB-I00 and by Spanish National Research Council (CSIC) Research Platform PTI-001. 
Y.B. acknowledges support from the Spanish Government via the project PID2021-126694NA-C22 (MCIU/AEI/FEDER, EU).
\end{acknowledgments}

\appendix

\section{Analytical solution to the first- and second-order NC for an SSH chain with \texorpdfstring{$2N-1$}{2N-1} sites}\label{sec:Appendix2NC}

We derive, step by step, the approximate gauge potentials $\mathcal{A}_\lambda^{(1)}$ starting from the base Hamiltonian [Eq.~\eqref{eq:ssh}] to retrieve the analytical expressions of $\alpha_1^{(1)}(t)$ from Eq.~\eqref{eq:gauge_exp} via action minimization. To ease the calculations, we recall that spinless creation and annihilation operators obey the anticommutation rules $\{c_i,c_j\}=0$ and $\{c^\dagger_i,c_j\}=\delta_{ij}$ with $\delta_{ij}$ equal to 1 if $i=j$ and 0 otherwise. Therefore, $[c^\dagger_ic_j,c^\dagger_kc_l]=c^\dagger_ic_l\delta_{jk}-c^\dagger_kc_j\delta_{il}$. We start from the reference Hamiltonian [Eq.~\eqref{eq:ssh}] with hopping parameters $t_1(t)=\lambda(t)$ and $t_2(t)=\Lambda-\lambda(t)$, where $\Lambda$ is some constant given by the chosen parametrization of the hoppings. Before computing the first-order NC~\eqref{eq:first_order_nested_commutator}, we need the following quantities:
\begin{wide}
\begin{align}\label{eq:1nc_general}
 \frac{[H_0,\partial_\lambda H_0]}{\Lambda}&=\left[ \sum_{j=1}^{N-1}c^\dagger_{2j+1}c_{2j-1} - \sum_{j=1}^{N-2}c^\dagger_{2j+2}c_{2j} \right]-\hc,\\ \label{eq:com_for_g1}
 \begin{split}
 \frac{[H_0,[H_0,\partial_\lambda H_0]]}{\Lambda}&=t_2\left[2\sum_{j=1}^{N-2}(c^\dagger_{2j+1}c_{2j}-c^\dagger_{2j+2}c_{2j})+c^\dagger_{2N-1}c_{2N-2}\right] \\
 &+t_1\left[\sum_{j=1}^{N-2}(2c^\dagger_{2j+3}c_{2j}-c^\dagger_{2j+2}c_{2j+1}-c^\dagger_{2j}c_{2j-1})-c^\dagger_{2N-2}c_{2N-3}\right]+\hc
 \end{split}
\end{align}
We can compute $G_1=\partial_\lambda H_0-i[H_0,\mathcal{A}_\lambda^{(1)}]$ for $2N-1$ sites due to Eq.~\eqref{eq:com_for_g1}. Grouping terms, one can derive
\begin{equation}\label{eq:g1}
 \begin{split}
  G_1&=\sum_{j=1}^{N-2}\left[(1+2\alpha^{(1)}_1\Lambda t_2)c^\dagger_{2j+1}c_{2j}-(1+2\alpha^{(1)}_1\Lambda t_1)c^\dagger_{2j+2}c_{2j+1}-2\alpha^{(1)}_1\Lambda t_2c^\dagger_{2j+2}c_{2j-1}+2\alpha^{(1)}_1\Lambda t_1c^\dagger_{2j+3}c_{2j}\right] \\
  &\qquad\qquad+(1+\alpha^{(1)}_1\Lambda t_2)c^\dagger_{2N-1}c_{2N-2}-(1+\alpha^{(1)}_1\Lambda t_1)c^\dagger_2c_1 + \hc
 \end{split}
\end{equation}
\end{wide}%
Following Wick's theorem~\cite{Sels_2017}, the only surviving terms of the action $S_1=\trace{G^2_1}$ will be proportional to the sum of squares of individual contributions in the previous equation. Therefore,
\begin{equation}\label{eq:s1}
\begin{split}
 \frac{S_1}{2}&=\sum_{i=1}^2\Big((N-2)\left[(1+2\alpha^{(1)}_1\Lambda t_i)^2+(2\alpha^{(1)}_1\Lambda t_i)^2\right]\\
 &\qquad+(1+\alpha^{(1)}_1\Lambda t_i)^2\Big).
\end{split}
\end{equation}
Minimization of the action $S_1$ solving $\partial S_1/\partial\alpha^{(1)}_1=0$ implies that
\begin{equation}
 \alpha_1^{(1)}=-\frac{2(N-2)+1}{[8(N-2)+1](t_1^2+t_2^2)}=-\frac{\mathcal{C}(N)}{t^2_1+t^2_2},
\end{equation}
with $\mathcal{C}(N)=[2(N-2)+1]/[8(N-2)+1]$. In particular, for five sites, $\mathcal{C}(3)=1/3$.

To derive the second-order NC we have to minimize the action $S_2=\trace{G_2^2}$. Apart from Eqs.~\eqref{eq:1nc_general} and~\eqref{eq:com_for_g1}, the following quantities are also needed:
\begin{wide}
\begin{align}
\begin{split}
 \label{eq:2nc_general}
 \frac{[H_0,[H_0,[H_0,\partial_\lambda H_0]]]}{\Lambda}&=t^2_2\left[4\sum_{j=1}^{N-2}(c^\dagger_{2j+1}c_{2j-1}-c^\dagger_{2j+2}c_{2j})+c^\dagger_{2N-1}c_{2N-3}\right] \\
 &+t^2_1\left[\sum_{j=1}^{N-2}(3c^\dagger_{2j+3}c_{2j+1}+c^\dagger_{2j+1}c_{2j-1}-4c^\dagger_{2j+2}c_{2j})+c^\dagger_{2N-1}c_{2N-3}\right] \\
 &+4t_1t_2\left[\sum_{j=1}^{N-3}(c^\dagger_{2j+3}c_{2j-1}-c^\dagger_{2j+4}c_{2j})+c^\dagger_{2N-1}c_{2N-5}\right]-\hc,
\end{split} \\ \label{eq:com_for_g2}
\begin{split}
 \frac{[H_0,[H_0,[H_0,[H_0,\partial_\lambda H_0]]]]}{\Lambda}&=t^3_2\left[8\sum_{j=1}^{N-2}(c^\dagger_{2j+1}c_{2j}-c^\dagger_{2j+2}c_{2j-1})+c^\dagger_{2N-1}c_{2N-2}\right]\\
 &+t^3_1\left[\sum_{j=1}^{N-2}(8c^\dagger_{2j+3}c_{2j}-c^\dagger_{2j}c_{2j-1}-7c^\dagger_{2j+2}c_{2j+1})-c^\dagger_{2N-2}c_{2N-3}\right] \\
 &+t_1t^2_2\left[\sum_{j=1}^{N-2}(9c^\dagger_{2j+3}c_{2j}-4c^\dagger_{2j}c_{2j-1}-4c^\dagger_{2j+2}c_{2j+1})-c^\dagger_{2N-2}c_{2N-3}\right] \\
 &+t^2_1t_2\left[\sum_{j=1}^{N-2}(5c^\dagger_{2j+1}c_{2j}-9c^\dagger_{2j+2}c_{2j-1})+4c^\dagger_{2N-1}c_{2N-2}\right] \\
 &+t_1t_2\sum_{j=1}^{N-3}[t_2(7c^\dagger_{2j+3}c_{2j}-8c^\dagger_{2j+4}c_{2j-1}) + t_1(8c^\dagger_{2j+5}c_{2j}+3c^\dagger_{2j+3}c_{2j+2}-7c^\dagger_{2j+4}c_{2j+1})]+\hc
\end{split}
\end{align}
Note that some of the latter terms of Eqs.~\eqref{eq:2nc_general} and~\eqref{eq:com_for_g2} only appear if $N\ge4$. Following the same strategy as in Eqs.~\eqref{eq:g1} and~\eqref{eq:s1}, $G_2=\partial_\lambda H_0-i[H_0, \mathcal{A}_\lambda^{(2)}]$ and $S_2=\trace{G_2^2}$ return
\begin{align}
 \begin{split}
  G_2 &=\sum_{j=1}^{N-3}\Big[2\Lambda \left(t_1[\alpha^{(2)}_1+4\alpha^{(2)}_2(t^2_1+2t^2_2)]c^\dagger_{2j+3}c_{2j}- t_2[\alpha^{(2)}_1+4\alpha^{(2)}_2(2t^2_1+t^2_2)]c^\dagger_{2j+4}c_{2j+1}+4\alpha^{(2)}_2t_1t_2[t_1c^\dagger_{2j+5}c_{2j}-t_2c^\dagger_{2j+4}c_{2j-1}]\right) \\
  &\qquad+\left(1+2\Lambda t_1[\alpha^{(2)}_1+4\alpha^{(2)}_2(t^2_1+t^2_2)]\right)c^\dagger_{2j+2}c_{2j+1}-\left(1+2\Lambda t_2[\alpha^{(2)}_1+4\alpha^{(2)}_2(t^2_1+t^2_2)]\right)c^\dagger_{2j+3}c_{2j+2}\Big] \\
  &-\Lambda t_2[2\alpha^{(2)}_1+\alpha^{(2)}_2(9t^2_1+8t^2_2)]c^\dagger_4c_1+\Lambda t_1[2\alpha^{(2)}_1+\alpha^{(2)}_2(8t^2_1+9t^2_2)]c^\dagger_{2N-1}c_{2N-4} \\
  &-\left(1+\Lambda t_1[\alpha^{(2)}_1+\alpha^{(2)}_2(t^2_1+4t^2_2)]\right)c^\dagger_2c_1+\left(1+\Lambda t_2[2\alpha^{(2)}_1+\alpha^{(2)}_2(5t^2_1+8t^2_2)]\right)c^\dagger_3c_2 \\
  &-\left(1+\Lambda t_1[2\alpha^{(2)}_1+\alpha^{(2)}_2(8t^2_1+5t^2_2)]\right)c^\dagger_{2N-2}c_{2N-3}+\left(1+\Lambda t_2[\alpha^{(2)}_1+\alpha^{(2)}_2(4t^2_1+t^2_2)]\right)c^\dagger_{2N-1}c_{2N-2}+\hc,
 \end{split} \\
 \begin{split}
  \frac{S_2}{2} &=(N-3)\Big[4\Lambda^2\left( t^2_1[\alpha^{(2)}_1+4\alpha^{(2)}_2(t^2_1+2t^2_2)]^2+t^2_2[\alpha^{(2)}_1+4\alpha^{(2)}_2(2t^2_1+t^2_2)]^2+[4\alpha^{(2)}_2t_1t_2]^2(t^2_1+t^2_2)\right) \\
  &\qquad+\left(1+2\Lambda t_1[\alpha^{(2)}_1+4\alpha^{(2)}_2(t^2_1+t^2_2)]\right)^2+\left(1+2\Lambda t_2[\alpha^{(2)}_1+4\alpha^{(2)}_2(t^2_1+t^2_2)]\right)^2\Big] \\
  &+\Lambda^2\left(t^2_2[2\alpha^{(2)}_1+\alpha^{(2)}_2(9t^2_1+8t^2_2)]^2+t^2_1[2\alpha^{(2)}_1+\alpha^{(2)}_2(8t^2_1+9t^2_2)]^2\right) +\left(1+\Lambda t_1[\alpha^{(2)}_1+\alpha^{(2)}_2(t^2_1+4t^2_2)]\right)^2 \\
  &+\left(1+\Lambda t_2[2\alpha^{(2)}_1+\alpha^{(2)}_2(5t^2_1+8t^2_2)]\right)^2 + \left(1+\Lambda t_1[2\alpha^{(2)}_1+\alpha^{(2)}_2(8t^2_1+5t^2_2)]\right)^2 +\left(1+\Lambda t_2[\alpha^{(2)}_1+\alpha^{(2)}_2(4t^2_1+t^2_2)]\right)^2.
 \end{split}
\end{align}
We now minimize the action $S_2$ by solving $\partial S_2/\partial\alpha^{(2)}_{1,2}=0$, yielding
\begin{align} \label{eq:a1_from_a2}
 \frac{\partial S_2}{\partial\alpha^{(2)}_1}=0\implies\alpha^{(2)}_1&=-\frac{\alpha^{(2)}_2\left([32(N-2)+1](t^4_1+t^4_2)+32[3(N-3)+2]t^2_1t^2_2\right)
 }{[8(N-2)+1](t^2_1+t^2_2)} - \frac{\mathcal{C}(N)}{t^2_1+t^2_2}, \\ \label{eq:a2_from_a1}
 \frac{\partial S_2}{\partial\alpha^{(2)}_2}=0\implies\alpha^{(2)}_2&=-\frac{[8(N-2)+1](t^2_1+t^2_2)+\alpha^{(2)}_1\left([32(N-2)+1](t^4_1+t^4_2)+32[3(N-3)+2]t^2_1t^2_2\right)}{[128(N-2)+1](t^6_1+t^6_2)+6[128(N-3)+59]t^2_1t^2_2(t^2_1+t^2_2)}.
\end{align}
We can solve $\alpha^{(2)}_1$ and $\alpha^{(2)}_2$ by solving the system of linear equations given by Eqs.~\eqref{eq:a1_from_a2} and~\eqref{eq:a2_from_a1}, returning
\begin{align}\label{eq:alpha12}
 \alpha^{(2)}_1&= -\frac{(t^2_1+t^2_2)\left(90(N-2)(t^4_1+t^4_2)+[2N(256N-1039)+1725]t^2_1t^2_2\right)}{72(N-2)(t^8_1+t^8_2)+[8N(128N-581)+4851]t^2_1t^2_2(t^4_1+t^4_2)+2[16N(32N-99)+193]t^4_1t^4_2}, \\ \label{eq:alpha22}
 \alpha^{(2)}_2 &= \frac{18(N-2)(t^4_1+t^4_2)+2[32N(N-4)+111]t^2_1t^2_2}{72(N-2)(t^8_1+t^8_2)+[8N(128N-581)+4851]t^2_1t^2_2(t^4_1+t^4_2)+2[16N(32N-99)+193]t^4_1t^4_2}.
\end{align}
\end{wide}%
In particular, for five sites we have that
\begin{align}
 \alpha^{(2)}_1&= -\frac{9(t^2_1+t^2_2)\left[10(t^4_1+t^4_2)+11t^2_1t^2_2\right]}{72(t^8_1+t^8_2)+123t^2_1t^2_2(t^4_1+t^4_2)+98t^4_1t^4_2}, \\
 \alpha^{(2)}_2 &= \frac{6\left[3(t^4_1+t^4_2)+5t^2_1t^2_2\right]}{72(t^8_1+t^8_2)+123t^2_1t^2_2(t^4_1+t^4_2)+98t^4_1t^4_2}.
\end{align}
In~\figlabel{fig:2nc}, the time-dependent Schrödinger equation is solved for an SSH chain of five sites, setting $\Omega\in[0.1,0.2,0.5,1]$ and $t_0=1$ without and with the addition of the second-order NC contribution (left and right columns, respectively) showcasing the robustness of the protocol against nonadiabatic effects.
\begin{figure*}[!tb]
\centering
\subfloat{%
\centering
\includegraphics{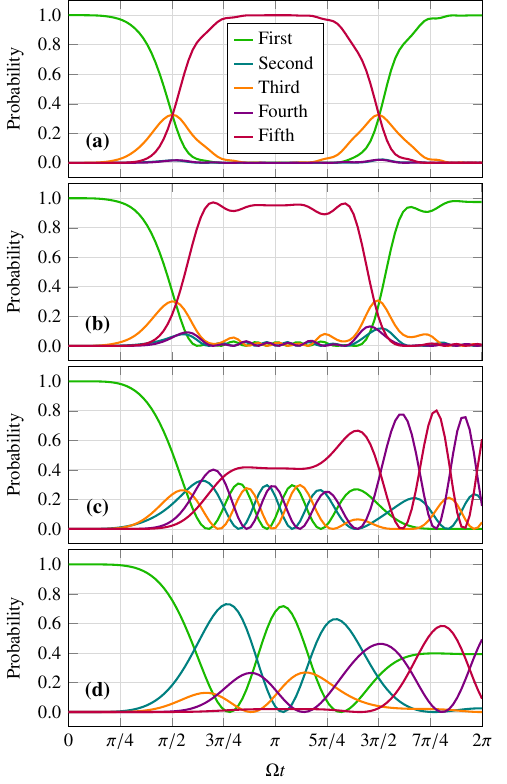}%
}\hspace{.5cm}%
\subfloat{%
\centering
\includegraphics{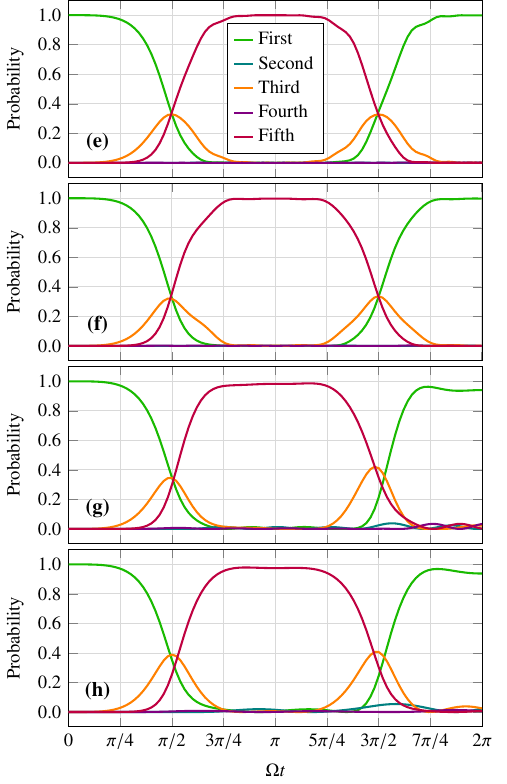}%
}%
\caption{On-site probability time evolution starting from $\ket{\Phi_0(0)}$ by solving the time-dependent SE without and adding the second order NC contribution (\textbf{(a)}-\textbf{(d)} and~\textbf{(e)}-\textbf{(h)}, respectively). From top to bottom on both subfigures: $\Omega$ is equal to $0.1$, $0.2$, $0.5$ and $1$.}\label{fig:2nc}
\end{figure*}

\section{Relation between the order of NC and their interactions involved} \label{sec:d-NC-2N-1}
 As shown in Eqs.~\eqref{eq:1nc_general} and~\eqref{eq:2nc_general}, the first-order NC introduces interactions among NNNs and the second-order NC introduces interactions between NNNs and fourth-nearest neighbors. This behavior extends to arbitrary order of NC.

 For the $d$th-order NC ($d<N-1$), interactions with the $2l$th nearest neighbors ($l\in[1,d]$) will appear. Considering terms such as $c^\dagger_{2(m+l)-r}c_{2m-r}$ with $2(m+l)-r<2N-1$ and $r\in\{0,1\}$, the $(d+1)$th-order NC will be given by
 \begin{equation}
 \begin{split}
   &[c^\dagger_{2j+p}c_{2j+p-1},[c^\dagger_{2k+q}c_{2k+q-1},c^\dagger_{2(m+l)-r}c_{2m-r}]]\\
   &\qquad=[c^\dagger_{2j+p}c_{2j+p-1},c^\dagger_{2(m+l)-r+1}c_{2m-r}\delta_{r+q,1}-c^\dagger_{2(m+l)-r}c_{2m-r-1}\delta_{rq}] \\
   &\qquad=c^\dagger_{2(m+l)-r}c_{2(m-1)-r}\delta_{r+p,1}\delta_{rq}+c^\dagger_{2(m+l+1)-r}c_{2m-r}\delta_{rp}\delta_{r+q,1} \\
   &\qquad -c^\dagger_{2(m+l)-r+1}c_{2m-r-1}(\delta_{r+p,1}\delta_{rq}+\delta_{rp}\delta_{r+q,1})
 \end{split}
 \end{equation}
 so, as expected, the difference between fermionic operator site-indices increases by two in comparison with the original ones coming from the term $c^\dagger_{2(m+l)-r}c_{2m-r}$, which translates to the fact that $2(l+1)$th nearest neighbors are now also considered. 
 
 In conclusion, for an SSH model with $2N-1$ sites, $\mathcal{A}_\lambda^{(N-1)}$ in Eq.~\eqref{eq:Hcd} is the minimum order of the adiabatic gauge-potential expansion needed to involve all possible interactions along the same sublattice.

\section{Pauli-basis decomposition of SSH model}\label{sec:pauli_decomposition}
 Here we present the general Pauli-basis decomposition of the SSH-model-padded Hamiltonian with $2N-1$ sites and its corresponding first-order NC, based on the work presented in Refs.~\cite{Liu2022digital, romero2023paulicomposer}. Moreover, we do the same calculation for the Rice-Mele model~\cite{asboth2016ashort}. Note that at least $n=\lceil\log_2(2N-1)\rceil$ qubits will be needed to encode such Hamiltonians.

We start from the regular Hamiltonian [Eq.~\eqref{eq:ssh}] and split it into two separate contributions: one proportional to $t_2$ ($H_0^{t_2}$) and the other to $t_1$ ($H_0^{t_1}$), such that $H_0=t_2H_0^{t_2}+t_1H_0^{t_1}$.

Let $N-1=\sum_{i}c_i2^i$ and $N=\sum_{i}d_i2^i$ with $c_i, d_i\in\{0,1\}$ $\forall i$, and $\mathcal{D}_x\coloneqq[1+(-1)^xZ]/2$. For $N$ odd, we have
\begin{equation}
  H_0^{t_2}(N)=\left[\sum_{i=0}^{n-3}c_{n-2-i}\left(\bigotimes_{j=0}^{i-1}\mathcal{D}_{c_{n-2-j}}\right)\mathcal{D}_0I^{\otimes n-2-i}\right]X.
\end{equation}
For $N$ even, we have that
\begin{equation}
 \begin{split}
  H_0^{t_2}(N)&=H_0^{t_2}(N-1)+\left[\bigotimes_{i=0}^{n-3}\mathcal{D}_{c_{n-2-i}}\right]\mathcal{D}_0X.
 \end{split}
\end{equation}

Following Ref.~\cite{Liu2022digital}, before starting with the $H_0^{t_1}$ decomposition, let $\sigma^s_{\{0,1\}}\coloneqq\{X,Y\}$ (skew-diagonal Pauli matrices) and $\gamma_m\coloneqq(-1)^{(\lfloor n_Y/2\rfloor+\tilde{n}_Y)\text{ mod }2}$ with $\tilde{n}_Y$ be the number of $Y$ matrices in the $(m-1)$-rightmost skew-diagonal matrices of the decomposition. Moreover, we define
\begin{equation}
 \mathcal{S}^{\text{\{e,o\}}}_m \coloneqq \sum_{\substack{i_1,\dots,i_m\\n_Y\text{ \{even,odd\}}}}\frac{\gamma_m}{2^{m-1}}\bigotimes_{k=1}^m\sigma^s_{i_k},
\end{equation}
which returns a summation of all possible Pauli strings of length $m$ only involving $X$ or $Y$ matrices, with $n_Y$ even or odd. Such terms return Hermitian contributions for the tridiagonal matrix entries present in the SSH model. 

Therefore, the $H_0^{t_1}$ decomposition becomes
\begin{equation}
\begin{split}\label{eq:h_0_t1_general}
 H_0^{t_1}(N)&=\sum_{m=2}^{n-2}\left[\sum_{i=0}^{n-m-1}d_{n-2-i}\left(\bigotimes_{j=0}^{i-1}\mathcal{D}_{d_{n-2-j}}\right)\mathcal{D}_0I^{\otimes n-m-i-1}\right]\mathcal{S}^\text{e}_m \\ 
 &+\frac{I+Z^{c_{n-3}+1}}{2}\mathcal{S}^\text{e}_{n-1} + \mathcal{S}^\text{e}_n,
\end{split}
\end{equation}
except for $N$ an exact power of $2$, where Eq.~\eqref{eq:h_0_t1_general} simplifies to
\begin{equation}
 H_0^{t_1}(N)=\sum_{m=2}^nI^{\otimes n-m}\mathcal{S}^\text{e}_m.
\end{equation}
Parametrizing $t_{1,2}(t)$ as in Eq.~\eqref{eq:hopping_cos}, we have that $\partial_\lambda t_{1,2}=\pm1$, implying that $\partial_\lambda H_0=H_0^{t_1}-H_0^{t_2}$. With this information, we can decompose in the Pauli basis the matrix related to NNN couplings, taking into account that it is Hermitian with purely imaginary entries: thus $n_Y$ must be odd for each Pauli string of the resulting decomposition.

If we only consider sublattice-A NNN interactions, the corresponding Pauli-basis decomposition becomes
\begin{wide}
\begin{equation}\label{eq:1nc_com_odd_general}
  \mathcal{K}(N) = \sum_{m=1}^{n-3}\left[\sum_{i=0}^{n-m-1}d_{n-2-i}\left(\bigotimes_{j=0}^{i-1}\mathcal{D}_{d_{n-2-j}}\right)\mathcal{D}_0I^{\otimes n-m-i-2}\right]\mathcal{S}^\text{o}_m\mathcal{D}_0 + \left(\frac{I+Z^{c_{n-3}+1}}{2}\mathcal{S}^\text{o}_{n-2} + \mathcal{S}^\text{o}_{n-1}\right)\mathcal{D}_0
\end{equation}
\end{wide}%
except for $N$ an exact power of $2$, where Eq.~\eqref{eq:1nc_com_odd_general} reduces to
\begin{equation}
 \mathcal{K}(N) = \sum_{m=1}^{n-1}I^{\otimes n-m-1}\mathcal{S}^\text{o}_m\mathcal{D}_0.
\end{equation}

For completeness, the Rice-Mele model can be decomposed as
\begin{equation}
 \begin{split}
  \frac{1}{\Delta}H_\Delta(N)&=\left[\sum_{i=0}^{n-2}c_{n-2-i}\left(\bigotimes_{i=0}^{i-1}\mathcal{D}_{c_{n-2-i}}\right)\mathcal{D}_0I^{\otimes n-2-i}\right]Z \\
  &+\left[\bigotimes_{i=0}^{n-2}\mathcal{D}_{c_{n-2-i}}\right]\mathcal{D}_0.
 \end{split}
\end{equation}

\section{Another hopping parametrization}

We take polynomial hopping parameters from Ref.~\cite{DAngelis_2020}, where
\begin{equation}\label{eq:hoppings_poly}
  t_1(t) = t_0P(t/T), \quad t_2(t) = t_0[1-P(t/T)],
\end{equation}
with $P(x)=2x^3-3x^2+1$, satisfying that $P(0)=1$ and $P(1)=0$. If we take $t_1(t)=\lambda(t)\implies\dot{\lambda}(t)=6t_0t(1-t/T)/T^2$ and $t_2(t)=\Lambda -\lambda(t)$ with $\Lambda=t_0$, our previously derived results for computing general first- and second-order nested commutators [Eqs.~\eqref{eq:1nc_general}, \eqref{eq:2nc_general}, and \eqref{eq:alpha12}, \eqref{eq:alpha22}] can be used; thus our technique can be implemented analogously as shown for the hopping parameters given by Eq.~\eqref{eq:hopping_cos} with $\Lambda=t_0$ instead of $2t_0$. 

Moreover, based on Refs.~\cite{DAngelis_2020, asboth2016ashort}, we can propose
\begin{equation}
 t_1(t)=t_0\cos\Omega t/2,\quad t_2(t)=t_0\sin\Omega t/2,
\end{equation}
with transfer time $T=\pi/\Omega$ and now setting $t_1(t)=\lambda(t)\implies \dot{\lambda}(t)=-\Omega t_2(t)/2$ and $t_2(t)=\sqrt{t^2_0-\lambda^2(t)}$ $\forall t\in[0,T]$.

In~\figlabel{fig:hops}, the time evolution of all the hopping parametrizations reviewed in this paper is plotted. Note that Eqs.~\eqref{eq:hopping_cos} and \eqref{eq:hoppings_poly} almost overlap along the transfer window $t\in[0,T]$.

\begin{figure}[!bt]
    \centering
    \includegraphics{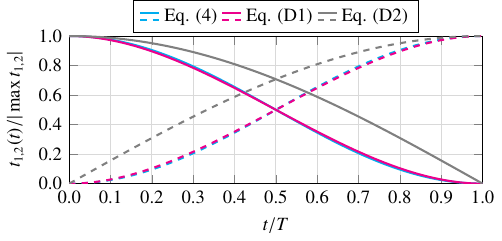}
    \caption{Hopping parametrizations addressed throughout the main text with $t_1$ ($t_2$) shown in solid (dashed) lines.}\label{fig:hops}
\end{figure}

\bibliography{bibfile}

\end{document}